# Dependency, Data and Decolonisation: A Framework for Decolonial Thinking in Collaborative AI Research

Dennis Reddyhoff, University of Sheffield, UK

This essay seeks to tie together thoughts on the political economy of academia, the inequities in access to the academic means of production and decolonial practice in data empowerment. To demonstrate this I will provide a brief analysis of the neo-colonial, extractive practices of the Western Academy, introduce concepts around decolonial AI practice and then use these to form an investigative framework. Using this framework, I present a brief case study of the AirQo project in Kampala, Uganda. The project aims to deploy a low-cost air pollution sensor network across the city, using machine learning methods to calibrate these sensors against reference instruments, providing high-quality air pollution data at a far lower cost.

When referring to the Western Academy, I refer to those researchers and institutions that have benefited from their position in the colonial hierarchy, and whose research practices have been extractive and exploitative. Depending on the area of study and time period, different groups could be considered to be a part of this bloc, but they are united in the windfalls they have seen from the imperialist history of which they are a part.

The Western academy has persisted as a part of the capitalist economic superstructure alongside colonialism and has evolved to reflect the neo-colonial, neo-liberal consensus of the Western economic bloc. As such, access to the academic means of production such as funding, knowledge and technology is inequitably distributed along colonial lines; these inequities inevitably lead to dependence on the West to provide this access to indigenous researchers. Thus, researchers and institutions outside of the academy are at a developmental disadvantage in establishing research autonomy. According to data from the UNESCO Institute for Statistics, the UK spends $43.8 billion on research and development (here we refer to purchasing power parity, or PPP$, allowing for a more direct comparison of spending), $11.2B of which is spent within the university sector and has 4,227 researchers per million inhabitants. Compare this to Uganda, which spends $114.5 million on R&D (0.2% GDP), of which $52 million is spent in the university sector with only 27 researchers per million inhabitants (UNESCO, 2016). These imbalances, brought about by colonialism and heightened by neo-colonialism create dependencies, whereby researchers from outside the academy rely on those within to access funding in addition to the ideas, education and technology that such funding provides.

> Decolonisation is a word much and unctuously used by imperialist spokesmen to describe the transfer of political control from colonialist to African sovereignty. The motive spring of colonialism, however, still controls the sovereignty. The young countries are still the providers of raw materials, the old of manufactured goods. The change in the economic relationship between the new sovereign states and the erstwhile masters is only one of

> form. Colonialism has achieved a new guise. It has become neo-colonialism, the last stage of imperialism … (Nkrumah, 1965)

While Kwame Nkrumah referred to the raw materials of production, when we observe research partnerships and collaboration between the West and those outside it through this neo-colonial lens, we see that many research partnerships also take on this extractive character, using researchers outside of the academy as data collectors while researchers within perform analytical roles (Chu et al., 2014). The value added at this analytical stage forms the basis of publication strategies and further grants. Meanwhile, the labour of the data collector goes largely unacknowledged, with less chance of dissemination and publication and little opportunity for academic and personal skill development. This lack of opportunity for development is compounded by inequities in access to journals, books and conferences creating what Syed Farid Alatas refers to as a "dependence on the media of ideas" (Alatas, 2003). Outside of the academy, large technology corporations and pseudo-monopolistic social media platforms such as Google, Facebook and Amazon extract, analyse and broker the data of all those with an online presence, generating huge profits for shareholders and investors in the West (Couldry and Medjias, 2018). A portion of these profits are then invested in R&D through research grants, offering 'Big Tech' a way in which to influence academic research policy through financial dependency. Indeed, 52% of tenure-track research faculty in the computer science departments of MIT, University of Toronto, Stanford and Berkeley have been directly funded by one of the Big Tech[1] companies (Abdalla and Abadalla, 2020). As such, there are power imbalances not just between those working within the academy and those outside it, but between institutions, research groups and academics whose aims best align with those of corporate funders and those that fall outside of their research interests.

Academic journals act as the gatekeepers of knowledge, hoarding research behind paywalls and subscriptions. With access to STEM journals costing upwards of $3000 per year per title on average, researchers and institutions with relatively low budgets face inequitable access to existing research. While the rising prevalence of open-access articles helps somewhat, the loss of revenue they represent to publishers has been offloaded onto authors, with open-access processing charges as high as $5000 for some journals (Bosch, 2019). Thus, researchers who have benefited from open access publication are barred from publishing under those same licenses by prohibitively high associated costs. Data, too, is a form of knowledge, one which can be used to educate, to explain and to advocate for change. However, closed-access proprietary hardware and software solutions act as neo-colonial instruments, building upon the funding and research advantages in the colonialist core and gating access to data behind high upfront investment and continued maintenance costs.

As researchers, we look to empower people through data. Pawelke and Canares present a vision of data empowerment that sets out to ensure that people have the ability to freely access data, the capacity to produce and use data relevant to them and to provide the freedom to exercise control over their personal data (Pawelke and Canares, 2018).

---

[1] Big Tech is defined as Google, Amazon, Facebook, Microsoft, Apple, Nvidia, Intel, IBM, Huawei, Samsung, Uber, Alibaba, Element AI and OpenAI in Abdulla and Abdulla, 2020.

Improved mobile and broadband internet infrastructure, Internet of Things technologies and machine learning methods present an opportunity to reduce the costs of data production and analysis. However, these technologies present new barriers in the costs of model computation and data storage. While these costs have continued to fall, we must also take into account the differences in R&D PPP available to indigenous researchers when considering the longevity and sustainability of research projects. Whilst data can empower people, if it cannot be accessed equitably due to cost, language or technological barriers then power imbalances arise between those countries which have the budget and existing technologies to collect, analyse and act upon data and those which cannot. We must take an additive-inclusive view of data production, looking to explore new and alternative approaches to the collection, analysis and presentation of empowering data (Mohamed et al., 2020).

When we look to decolonise our research, we must seek to turn an extractive relationship into an equitable one. Research partnerships must engage those from outside of the Western academy, putting their research needs at the centre of the project and ensuring that knowledge flows outward, seeking to provide education, skill development and a platform to develop research autonomy for the individual and research sovereignty for the institution. With this in mind, I have produced a list of questions to investigate the research project in Kampala, which I also hope will provide a lens through which we can all examine our past, current and future collaborations and better assess their neo-colonial aspects.

- Who is setting the research agenda, who is the PI?

  The project was started by Prof. Engineer Bainomughisa and Dr. Michael Smith at Makrere University. The project was continued by Prof. Bainomughisa, who now acts as PI and Project Lead with Dr. Smith acting as Co-Investigator at Sheffield University.

- Who is funding the research?

  The AirQo project is funded by Google as part of their AI Impact Grant. A separate related grant is funded by the EPSRC.

- Does it integrate people from outside the Western academy?

  Yes, over 80% of the team are Ugandan, based at the university or employed by AirQo. The project also hires local motorbike taxis to transport calibration sensors around the city.

- What authorship and dissemination prospects are there for local authors?

  Indigenous researchers are encouraged to disseminate their work in the form of presentations and conference/technical papers. Technical and mentoring support is provided by researchers in Sheffield and time is specifically allocated to these aspects in Research Associate contracts.

- How does the project become self-sustaining?

    The project has taken on a commercial aspect in order to sustain its research autonomy. While the positives and negatives of the commercialization of technology are rightly debated, it is the decision of the project leaders how best to seek compensation for their labour.

- Why is this data empowering? Who does it empower?

    The data will allow the public in Kampala to access accurate data on air pollution around their homes, schools and workplaces. The project is working with the city authorities, providing empirical data with which to shape their policies around air pollution,

- How is machine learning used to improve equity in access to this data?

    Gaussian Process methodologies are being employed to calibrate low-cost sensors against higher quality but considerably more expensive reference sensors. These methods seek to account for the inherent inaccuracies and drawbacks that the use of low-cost hardware presents. We are also seeking to use streaming methods in order to reduce computation and data costs to ensure long term viability after the initial funding period.

- Will developed hardware, software and research be open source/access?

    Air quality data is available through a free to use app. Unfortunately, in this case the raw sensor data is of little use without calibration.

- How will data access be maintained? How are ongoing costs minimised?

    Low-cost cloud computing solutions are being used by the team to build the required infrastructure for data collection and storage. Current efforts are underway to minimise these costs through better architectural and machine learning practices.

The historical material study of the Western Academy lays bare the extractive relationship it maintains with those outside its core and how the resulting imbalances lead to academic dependencies. While data, machine learning and the computer sciences have the potential to empower, they can also reproduce colonial power relations if we do not view our relationships through the neo-colonial lens. We must also now look towards an abolitionist analysis of academia as a whole, and what reforms we can demand and organise around without being reformist (Critical Resistance, 2019). While working toward decolonisation of our practices is commendable and worthwhile, the master's house will never be toppled with the master's tools, and so decolonial research practices cannot themselves decolonise the academy. In continuing to push toward decolonisation in research, we allow ourselves to better understand the conflicts and contradictions that academia contains and familiarise ourselves with ways to combat them. We must all continue to learn, write, teach and preach a decolonial line in our research, teaching and lives.